\documentclass{article}

\usepackage{arxiv}

\usepackage{graphicx}
\usepackage{listings}

\usepackage[utf8]{inputenc} % allow utf-8 input
\usepackage[T1]{fontenc}    % use 8-bit T1 fonts
\usepackage{hyperref}       % hyperlinks
\usepackage{url}            % simple URL typesetting
\usepackage{booktabs}       % professional-quality tables
\usepackage{amsfonts}       % blackboard math symbols
\usepackage{nicefrac}       % compact symbols for 1/2, etc.
\usepackage{microtype}      % microtypography
\usepackage{lipsum}

\title{Literature Review and Implementation Overview: High Performance Computing with Graphics Processing Units for Classroom and Research Use}

\author{
 Nathan George \\  % can use this right after name for more info \thanks{}
  Department of Data Sciences\\
  Regis University\\
  Denver, CO 80221 \\
  \texttt{ngeorge@regis.edu} \\
}

\begin{document}
\maketitle

\begin{abstract}
In this report, I discuss the history and current state of GPU HPC systems.  Although high-power GPUs have only existed a short time, they have found rapid adoption in deep learning applications.  I also discuss an implementation of a commodity-hardware NVIDIA GPU HPC cluster for deep learning research and academic teaching use.
\end{abstract}

% keywords can be removed
\keywords{GPU \and Neural Networks \and Deep Learning \and HPC}

\section{Introduction, Background, and GPU HPC History}

High performance computing (HPC) is typically characterized by large amounts of memory and processing power.  HPC, sometimes also called supercomputing, has been around since the 1960s with the introduction of the CDC STAR-100, and continues to push the limits of computing power and capabilities for large-scale problems \cite{hill_readings_1999, loshin_high_2014}.  However, use of graphics processing unit (GPU) in HPC supercomputers has only started in the mid to late 2000s \cite{showerman, fan_gpu_2004}.  Although graphics processing chips have been around since the 1970s, GPUs were not widely used for computations until the 2000s.  During the early 2000s, GPU clusters began to appear for HPC applications.  Most of these clusters were designed to run large calculations requiring vast computing power, and many clusters are still designed for that purpose  \cite{coates_deep_nodate}.  

\begin{figure}
\centering
\includegraphics[width=0.5\textwidth]{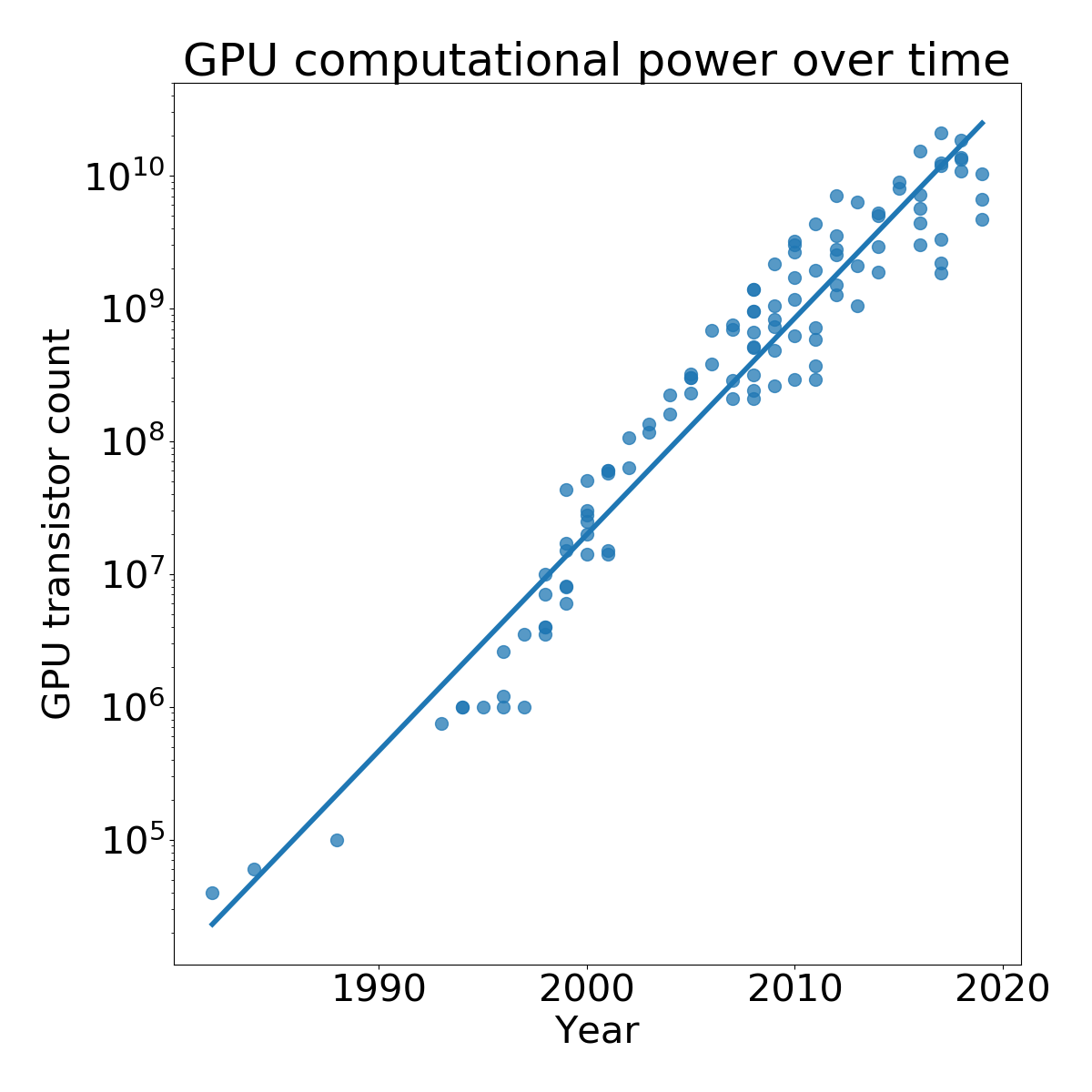}
\caption{A logarithmic plot of GPU transistor count (proportional to computational power) versus time shows Moore's Law holds true for GPUs \cite{george_nategeorgegpu_charts_2019}.}
\label{fig:gpu_moores}
\end{figure}

GPUs have been increasingly used for computations due to their commodification, following Moore's Law (demonstrated in Figure~\ref{fig:gpu_moores}), and usage in specific applications like neural networks.  Although server-grade GPUs can be used in clusters, commodity-grade GPUs are much more cost-effective.  A similar amount of computing power with commodity hardware can be obtained for roughly a third of the cost of server-grade hardware.  In 2018 NVIDIA suddenly forced businesses to replace commodity GPUs with their server-grade GPUs in what appeared to be primarily motivated by a desire to increase earnings, but may have been related to warranty issues as well \cite{nvidia_register_article}.  However, commodity hardware still proves to be useful for GPU clusters \cite{rohr_energy-efficient_2014, geveler_icarus_2017, maruyama_high-performance_2010}, especially in academic settings where the NVIDIA EULA does not seem to apply.  Several studies have examined the performance of commodity \cite{goddeke_exploring_2007, jeon_analysis_2019} and non-commodity \cite{debardeleben_gpu_2014} GPU hardware for various calculations, and generally found commodity hardware to be suitable for use in GPU clusters.  Although NVIDIA's legal definitions in their EULA are intentionally vague, it seems that using commodity NVIDIA GPUs and the associated NVIDIA drivers/software is allowed for smaller academic uses such as our use-case \cite{noauthor_nvidia_nodate}.

Although some guidelines exist for GPU clusters \cite{kindratenko_gpu_2009} and openHPC has ``recipes'' which are instructions for installing SLURM on a CentOS or SUSE cluster, there is no good step-by-step documentation for creating a commodity GPU cluster from scratch using Ubuntu Linux.  Ubuntu is currently one of the top-most used Linux distributions for both personal and server use and has a vibrant community as well as support, making Ubuntu a good choice for use as a Linux system.  One drawback of Ubuntu is it is frequently updated and may not be as stable as other Linux OS's such as SUSE and RHEL; even the long-term stability (LTS) version of Ubuntu has new versions released at least yearly, and is only officially supported for 3 years.  In this paper, I will talk about creating GPU clusters for multi-user utilization, where the goal is to maximize utility of commodity GPU hardware for end-users running neural network calculations.  I will also discuss implementation of a GPU HPC cluster at Regis University.

\section{Implementation}
\label{sec:Implementation}

\subsection{System Requirements}

Many GPU clusters are used to combine the power of several GPUs in order to complete a single calculation, or training of a single neural network.  Others have used these systems for specific problems such as protein folding, image analysis, and training large neural networks \cite{abdul-wahid_folding_2012, andrade_efficient_2014, yu_deep_2016}.  In our particular case, we required a system which uses a limited number of GPUs (6) to serve dozens of students and faculty.  Additionally, students will often use the system simultaneously during classes or peak hours, so we cannot simply assign one GPU per person as other institutions with dozens of GPUS may be able to.  As is common, we use Python neural network libraries (e.g. TensorFlow/Keras, PyTorch) along with NVIDIA's CUDA drivers, which completely take over a GPU upon running neural network training in Python, so that only one GPU can be used by one person at a time.  Another requirement is to easily create and manage large numbers of user credentials for the system.

\subsection{Hardware}
The instructions provided with this study used NVIDIA GPU cards \cite{george_nategeorge/slurm_gpu_ubuntu_2019}.  We are using commodity GPUs, which seems to be acceptable for use in small-scale labs (not datacenters) as per NVIDIA's EULA.  For current deep learning applications, it's wise to have as much GB of GPU RAM as possible.  Current commodity cards such as NVIDIA's 1080 Ti tend to be around 11GB per GPU card.  AMD GPUs should also be usable, although fewer deep learning libraries may be easily utilized with AMD GPUs.  For example, TensorFlow, PyTorch, and MXNet seem easy enough to use with AMD GPUs \cite{rocm_dl_libs, noauthor_rocmsoftwareplatform/mxnet_2019}, but other libraries such as Caffe appear to only have experimental support \cite{noauthor_bvlc/caffe_nodate}.

For maximum potential of the system, motherboards with four 16-lane PCIe slots can be used.  It is important to ensure the CPU(s) also provide enough PCI lanes for all 4 GPUs to fully use 16 PCI lanes per GPU, which maximizes data transfer speeds.  For Intel processors, it is likely that 2 CPUs are needed for 4 GPUs to fully utilize all PCI lanes with 4 GPUs and a M.2 SSD hard drive (which also uses PCIe lanes).   Some of AMD's processors, such as the EPYC line, currently support 64 or more PCI lanes.

The amount of RAM for the system can vary, but a minimum of 32GB per GPU card is recommended.  Some high-end motherboards can currently support up to 512GB of RAM.

Hard drive space is dependent on the use case, but for classroom use, a few hundred GBs per student is sufficient as a minimum.  Large datasets (100GB+) can be stored in a common location, and student-specific data will be relatively small compared with any training data.  Using LVM (Logical Volume Management) within Linux is a good idea for HDD storage so drives can be extended as more HDDs are added.

The power supply should be able to accommodate the GPUs as well as other components.  A currently common NVIDIA GPU, the 1080 Ti, uses around 250-300W of power at it's peak.  With four of these cards and other components, it is best to use a 1600W power supply (which is the largest commodity PSU currently available).

\subsubsection{Price comparison of server vs commodity hardware}
As has been the case for several years, commodity hardware has much lower prices than server-grade hardware.  A comparable 4-GPU setup as a pre-built server-grade system would cost roughly \$22,600, while a commodity-hardware based system would be around \$7,800 (see Appendix~A for more information).  This results in the server-grade hardware costing roughly 3 times more than the commodity hardware.  However, the server-grade system may be more reliable, come with more support, and have a better warranty than commodity hardware.

\subsection{Software}
\subsubsection{Job Management Software Options}

There are many software options for creating a GPU HPC cluster; software available for user and resource management with GPUs includes CUDA wrapper \cite{noauthor_cuda_wrapper_nodate} developed at the University of Illinios at Urbana-Champaign, SLURM \cite{yoo_slurm:_2003} (adapted for GPUs in 2014 \cite{iserte_slurm_2014, soner_extending_nodate}), OpenHPC \cite{openhpc}, Rocks \cite{noauthor_rocksclusters_nodate}, and many others \cite{noauthor_cluster_2011, noauthor_comparison_2019}.

A few proprietary/non-free cluster management solutions exist, including Bright Cluster Manager, StackIQ Boss, IBM Spectrum LSF, PBS Professional, TORQUE, and Grid Engine.  However, budgetary constraints may prohibit usage of these solutions.  Free and open-source solutions include SLURM, OpenHPC, and Rocks.  The downside of the open-source solutions is an increased amount of work to install and maintain the cluster.  Other solutions for aspects of GPU cluster management exist as well, such as Themis, which is a GPU workload scheduler \cite{mahajan_themis:_2019}.

One set of instructions exists for creating such clusters: OpenHPC's recipes \cite{openhpc}.  These include instructions for CentOS and Suse Linux Enterprise Server.  However, if one prefers Ubuntu for software reasons or familiarity with the Debian flavor of Linux, one can use instructions from the author of this paper \cite{george_nategeorge/slurm_gpu_ubuntu_2019}.

\subsubsection{Queuing system}
Of the open-source options, it appears SLURM is the best option for a job queuing management software.  SLURM is actively maintained (unlike Rocks), and is relatively easy to install and use.  Configuration of SLURM is made easier with auto-detection scripts for configurations, as well as sample configuration files \cite{george_nategeorge/slurm_gpu_ubuntu_2019}.

\subsubsection{OS configuration}

A few important requirements are necessary to have an HPC cluster run smoothly and be easily manageable:

\begin{itemize}
    \item synchronized GIDs/UIDs (group IDs and user IDs)
    \item synchronized system clocks
    \item a shared filesystem such as NFS (Network File System)
\end{itemize}

For SLURM specifically, we also require:
\begin{itemize}
    \item passwordless SSH between the master and worker nodes
    \item munge for authentication across the cluster
\end{itemize}

In our case, FreeIPA was found to be a convenient solution for syncing the GIDs/UIDs (which uses LDAP).  This uses chronyc to synchronize system clocks as well.

\subsubsection{Deep learning libraries}
Since deep learning can be easily done in Python (especially for teaching), we use common Python libraries to run neural networks.  After installing NVIDIA base GPU drivers on Ubuntu, the easiest way to install Python deep learning libraries is with the Anaconda Python distribution.  This takes care of installing NVIDIA deep learning software like CUDA and CuDNN, which can be time consuming to install manually.

\subsubsection{User Management}
Cluster user management seems to be something not widely discussed or published. For an academic use-case, we would like to be able to add many users at once, making sure their accounts expire at a predetermined date.  Creation of users is accomplished with a custom Bash script, which uses a CSV file to create both the Ubuntu OS users with FreeIPA, and the SLURM users with the same username \cite{george_nategeorge/slurm_gpu_ubuntu_2019}.  Storage quotas are set for students using the base Ubuntu OS tools, and filesystem permissions are set so as to prevent students from seeing the contents of other students' home folders.  SLURM user management also allows limits for users; we set limits for students as to how many jobs they can submit, how many concurrent jobs they can run at once, and how long their jobs can run.  This is to ensure all students will be able to use the cluster even during a study ``rush hour''.  We also set quality of service (QOS) levels for different users so that some users have a higher priority than others.  This results in some users having their jobs run more quickly if their QOS is a higher value.  For example, we give faculty a slightly higher priority than students for running jobs.

One difficulty with user management is that SLURM users are separate from OS users.  Of course it's easy enough to create both user accounts for a single user at once, but SLURM user accounts do not have the option to auto-expire.  Another small Bash script was written which runs daily to check if any users have expired from the Ubuntu UID list, and if so, the user is removed from SLURM \cite{george_nategeorge/slurm_gpu_ubuntu_2019}.

\subsubsection{Running deep learning jobs on the cluster}

Running jobs on the cluster is simple enough, although when something goes wrong it can be difficult to ascertain why.  Running code files like \texttt{job.py} can be run with the \texttt{srun} command:

\texttt{srun /home/msds/anaconda3/bin/python /storage/<username>/job.py}

where username is the Linux username such as \texttt{ngeorge}.  This will automatically put the job into a queue.  Note we specificy the exact path to the Python executable we want to use to avoid any ambiguity.  Rather than use \texttt{srun}, however, it is better practice to use a job file with \texttt{sbatch} since we can specify requirements for our job to run.  An example job file is:

\begin{lstlisting}[language=bash]
#!/bin/bash
#SBATCH -N 1      # nodes requested
#SBATCH --job-name=test
#SBATCH --output=/storage/<username>/test.out
#SBATCH --error=/storage/<username>/test.err
#SBATCH --time=2-00:00
#SBATCH --mem=36000
#SBATCH --qos=normal
#SBATCH --gres=gpu:1
srun /home/msds/anaconda3/bin/python /storage/<username>/tf_test.py
\end{lstlisting}

This can be submitted to the job queue with \texttt{sbatch job.job}, if the filename of the job file is \texttt{job.job}.  Each line preceded by \texttt{\#SBATCH} is an option for the \texttt{sbatch} command, with details listed in the \texttt{sbatch} documentation \cite{slurm_sbatch_docs}.

\section{Monitoring and debugging}

Once a job is submitted, it can be monitored through SLURM and NVIDIA software.  With SLURM, the \texttt{sacct} command can be useful.  This shows which jobs are running for the current account user.  The command can be customized to show other fields \cite{slurm_sacct_docs}, for example:

\texttt{sacct --format=jobid,jobname,state,exitcode,user,account}

Another useful SLURM command is \texttt{sinfo}, which shows the status of the nodes in the cluster.  Monitoring the hardware itself can be done with a few tools: \texttt{nvidia-smi} and \texttt{htop}.  \texttt{nvidia-smi} will show the utilization of NVIDIA GPUs on a machine, and \texttt{htop} will show RAM and CPU utilization.

\section{Conclusion}
HPC GPU clusters have been around since the early 2000s, but little exists in the way of instructions and overviews of creating HPC GPU clusters for shared use.  Here, we discussed how an HPC GPU cluster can be created using NVIDIA commodity GPUs with commodity hardware, free and open-source software including Ubuntu and SLURM, and users for the cluster can be managed with custom Bash scripts.  This solution works for smaller installations where sharing GPUs amongst multiple users is a requirement.  For larger clusters, one should use server-grade NVIDIA GPUs as per their EULA (or risk being sued), or use AMD GPUs.

\section{Appendices}

\subsection{A: Server versus commodity hardware comparison}

Similar hardware setups were priced between commodity-based hardware and server-grade for a comparison of cost, resulting in a total cost of \$22,615 for the server-grade hardware \cite{thinkmate}, and a cost of \$7845.01 for the commodity hardware \cite{newegg_wishlist}.  This results in server-grade hardware roughly being a factor of 3 times more expensive than the commodity hardware.  One large factor in the cost was the GPUs, which were roughly double the cost for the server-grade hardware compared with commodity hardware.

Components were chosen in order to make the hardware setups comparable, but also efficient.  For example, no Intel CPU currently exists which has more than 64 PCI lanes, but AMD's EPYC Rome CPU series has PCI lanes up to 128.  Since each GPU requires 16 PCI lanes for maximum data transfer efficiency, and the M.2 SSD also requires PCI lanes, more than 64 PCI lanes are required for optimum performance.  Consequently, an AMD processor was chosen for the commodity setup.  AMD processors were not available for server-grade builds, so Intel was chosen.  For the server-grade hardware vs commodity hardware, comparable GPUs were chosen based on CUDA cores and GPU RAM.  The server-grade GPU (NVIDIA® Quadro® RTX 5000 16.0GB GDDR6 (4xDP, 1xVirtualLink)) has slightly more RAM but fewer CUDA cores than the commodity NVIDIA 1080 Ti GPU.

Server-grade prices were obtained from thinkmate.com on April 13th, 2020 \cite{thinkmate}.  The configuration (model XT24-24S1-8GPU) chosen was as follows:
 
\begin{center}
\def\arraystretch{1.5}
 \begin{tabular}{|p{2.3cm}|p{11cm}|l|} 
 \hline
 Component & Description & Number\\ [0.5ex] 
 \hline\hline
 Motherboard and PSU & Intel\textsuperscript{\textregistered} C622 Chipset - 4U GPU Server - 24x SATA - Dual 10-Gigabit Ethernet - 2000W (2+2) Redundant Power Supply & 1\\ 
 \hline
Processor & Intel® Xeon® Gold 5215 Processor 10-Core 2.5GHz 14MB Cache (85W) & 2 \\
 \hline
 RAM & 8GB PC4-21400 2933MHz DDR4 ECC RDIMM & 12 \\
 \hline
 Primary Hard Drive & 1.0TB Intel® SSD DC P4101 Series M.2 PCIe 3.1 x4 NVMe Solid State Drive & 1 \\
 \hline
 Data Hard Drives & 3.84TB Micron 5200 PRO Series 2.5" SATA 6.0Gb/s Solid State Drive & 5 \\ 
 \hline
 GPUs & NVIDIA® Quadro® RTX 5000 16.0GB GDDR6 (4xDP, 1xVirtualLink)& 4 \\ [1ex]
 \hline
\end{tabular}
\end{center}

\begin{center}
\def\arraystretch{1.5}
 \begin{tabular}{|l|l|} 
 \hline
 Component & Total Price Increase from Base (\$)\\ [0.5ex] 
 \hline\hline
 Motherboard and PSU & 0\\ 
 \hline
Processor & 2400 \\
 \hline
 RAM & 0\\
 \hline
 Primary Hard Drive & 49 \\
 \hline
 Data Hard Drives & 3675 \\ 
 \hline
 GPUs & 9,196 \\ [1ex]
 \hline
\end{tabular}
\end{center}

The overall price was \$22,615 before taxes (base price before customization and adding GPUs was \$7,474).

Commodity-grade hardware prices were obtained from newegg.com on April 13th, 2020 \cite{newegg_wishlist}.  The commodity machine cost was \$7845.01 before taxes.  The specifications follow:

\begin{center}
\def\arraystretch{1.5}
 \begin{tabular}{|p{2.3cm}|p{11cm}|l|} 
 \hline
 Component & Description & Number\\ [0.5ex] 
 \hline\hline
 Motherboard & AsRock Rack EPYCD8-2T ATX Server Motherboard AMD EPYC 7002/7001 (Naples/Rome) Series SP3 LGA4094 Dual 10 GLAN & 1\\ 
 \hline
 Power Supply & EVGA SuperNOVA 1600 G2 120-G2-1600-X1 80+ GOLD 1600W Fully Modular Includes FREE Power On Self Tester Power Supply & 1 \\
 \hline
Processor &  AMD EPYC Rome 7302P 16-Core 3.0 GHz (3.3 GHz Max Boost) Socket SP3 155W 100-100000049WOF Server Processor & 1 \\
\hline
CPU cooling & DEEPCOOL Castle 240EX, Addressable RGB AIO Liquid CPU Cooler, Anti-Leak Technology Inside, Cable Controller and 5V ADD RGB 3-Pin Motherboard Control, TR4/AM4 Supported & 1 \\
 \hline
 RAM & CORSAIR Vengeance LPX 64GB (2 x 32GB) 288-Pin DDR4 SDRAM DDR4 2400 (PC4 19200) Desktop Memory Model CMK64GX4M2A2400C16 & 2 \\
 \hline
 Primary Hard Drive & Intel 660p Series M.2 2280 1TB PCIe NVMe 3.0 x4 3D2, QLC Internal Solid State Drive (SSD) & 1 \\
 \hline
 Data Hard Drives & WD Red 10TB NAS Hard Disk Drive - 5400 RPM Class SATA 6Gb/s 256MB Cache 3.5 Inch - WD100EFAX & 2 \\ 
 \hline
 GPUs & EVGA GeForce GTX 1080 Ti SC2 GAMING, 11G-P4-6593-KR, 11GB GDDR5X & 4 \\ 
 \hline
 Case & LIAN LI O11 Dynamic XL ROG certificated -Black color ---Tempered Glass on the Front, and Left Side. E-ATX ,ATX Full Tower Gaming Computer Case---O11D XL-X & 1 \\
 [1ex]
 \hline
\end{tabular}
\end{center}

\begin{center}
\def\arraystretch{1.5}
 \begin{tabular}{|l|l|} 
 \hline
 Component & Cost (\$)\\ [0.5ex] 
 \hline\hline
 Motherboard & 487.00 \\ 
 \hline
 Power Supply & 461.63 \\
 \hline
Processor & 890.45 \\
\hline
CPU cooling & 114.99 \\
 \hline
 RAM & 599.98\\
 \hline
 Primary Hard Drive & 124.99 \\
 \hline
 Data Hard Drives & 569.98\\ 
 \hline
 GPUs & 4,396 \\ 
 \hline
 Case & 199.99 \\
 [1ex]
 \hline
\end{tabular}
\end{center}

\bibliographystyle{unsrt}


\end{document}